\documentclass[showpacs,aps,amsmath,amssymb,twocolumn]{revtex4}
\usepackage[latin1]{inputenc}
\usepackage{graphicx}
\usepackage{bm}
\usepackage{array}
\usepackage{braket}
\usepackage{microtype}
\usepackage{wasysym}
\renewcommand{\baselinestretch}{1.0}
\begin{document}

\renewcommand{\baselinestretch}{1.0}

\title{Proposal for a Direct Measurement of the von Neumann Entropy and the Relative Entropy of Coherence}

\author{Bert\'{u}lio de Lima Bernardo$^{1,2}$}

\affiliation{$^{1}$Departamento de F\'{\i}sica, Universidade Federal da Para\'{\i}ba, 58051-900 Jo\~ao Pessoa, PB, Brazil\\
$^{2}$Departamento de F\'{\i}sica, Universidade Federal de Campina Grande, Caixa Postal 10071, 58109-970 Campina Grande-PB, Brazil}

\email{bertulio.fisica@gmail.com}

\begin{abstract}

Quantum coherence is the most distinguished signature of quantum mechanics, also recognized to be an essential resource for many promising quantum technologies, playing a central role in phenomena related to quantum information science, quantum thermodynamics and quantum biology, just to mention few examples. However, the resource theory of coherence is in the preliminary stage of its development, being still limited to the study of the manipulation and quantification of coherence, mostly from a theoretical viewpoint. Here, we propose an experimental method to directly measure the relative entropy of coherence, which according to the resource theory is one of the main quantifiers of coherence. To achieve this aim, we discuss a procedure to measure the von Neumann entropy of a generic quantum state directly in terms of the Shannon entropy of the probability distribution of outcomes in a defined measurement basis. In both cases, by direct we mean that tomographic methods are not required. Two quantum-optical applications of our method are discussed in order to give support to our predictions.

\end{abstract}

\maketitle


\section{introduction}

Coherence, along with quantum entanglement and the measurement process, is one of the characteristic attributes that distinguishes the quantum theory from the classical realm. Specifically, the concept of quantum entanglement only emerges when one extends the notion of quantum coherence to the scope of many-body physical systems, which leads us to understand that the latter concept is even more fundamental than the former \cite{streltsov,horodecki}. In this regard, quantum coherence is now viewed a key element behind a great number of quantum information processing tasks \cite{nielsen}, which differently could never be realized by means of classical methods. It has also been discovered that quantum coherence plays an essential role in the study of quantum thermodynamics \cite{lost1,lost2,kam}, quantum metrology \cite{gio1,gio2}, quantum biology \cite{abbott}, spin models \cite{karpat}, transport theory \cite{deve}, and many other researches related to quantum theory \cite{chen,cheng,hu,mondal,bert,bert2}.    

Although the long-standing investigations of the properties of coherence, which has been realized prominently in the scenario of quantum-optical systems since the inception of quantum theory \cite{bohr,feynman}, only recently a theory of quantum coherence as a resource was presented in the seminal work by Baumgratz {\it et al.} \cite{baum}. This work has triggered renewed interest on the subject in a variety of quantum mechanical and nanoscale systems \cite{giro,chanda,yao}. In such a resource-theoretical framework, two fundamental starting points are the definition of incoherent operations, and the choice of proper measures of coherence. In this context, Baumgratz {\it et al.} \cite{baum} defined incoherent states those which are diagonal in a given (incoherent) basis $\{\ket{i}\}$, that is, all states of the type $\hat{\rho} = \sum_{i} p_{i} \ket{i}\bra{i}$. The incoherent operations $\Lambda [\hat{\rho}]$, i.e., those which cannot create coherence, specified by a set
of Kraus operators $\{ \hat{K}_{n} \}$, with $\sum_{n} \hat{K}^{\dagger}_{n} \hat{K}_{n} = I$, are required to satisfy the map $\Lambda [\mathcal{I}] = \hat{K}_{n} \mathcal{I} \hat{K}^{\dagger}_{n} \subseteq \mathcal{I}$, where $\mathcal{I}$ is the set of incoherent states. Moreover, in a Hilbert space of dimension $d$, the maximally coherent state is $\ket{\Phi_{d}}= \sqrt{1/d} \sum_{i=1}^{d} \ket{i}$, where the state $\ket{\Phi_{2}}$ is defined to contain the unit of coherence (the coherence bit, or {\it cobit}). Following these ideas, significant progress has been achieved within the domain of the resource theory. Many of these works are devoted to investigate the properties of the coherence measures \cite{yuan,streltsov2}, and the search for new coherence measure candidates \cite{streltsov3,winter}.

The rigorous formalism of the resource theory of coherence requires that a proper measure of coherence $C(\hat{\rho})$ of a quantum state $\hat{\rho}$ should fulfill a set of criteria \cite{streltsov}: (i) Non-negativity: $C(\hat{\rho}) 	\ge 0$, with the equality applied to incoherent states. (ii) Monotonicity: incoherent operations does not increase coherence, i.e., $C(\hat{\rho}) \ge C(\Lambda [\hat{\rho}])$. (iii) Strong monotonicity: on average, $C(\hat{\rho})$ does not increase upon selective incoherent operations, $C(\hat{\rho}) 	\ge \sum_{n} p_{n} C(\hat{\rho}_{n})$, with probabilities $p_{n} = Tr [\hat{K}_{n} \hat{\rho} \hat{K}^{\dagger}_{n}]$ and postmeasurement states $\hat{\rho}_{n} = \hat{K}_{n} \hat{\rho} \hat{K}^{\dagger}_{n}/p_{n}$. (iv) Convexity: $C(\hat{\rho})$ is a convex function, $\sum_{n} p_{n} C(\hat{\rho}_{n}) \ge C\left(\sum_{n} p_{n} \hat{\rho}_{n} \right)$. However, by following the resource theory of quantum entanglement \cite{horodecki}, for a coherence quantifier to be considered a rigorous measure of coherence, it also has to satisfy two other more stringent conditions: (v) Uniqueness for pure states: given a pure state $\ket{\psi}$, it is necessary that $C(\ket{\psi}\bra{\psi}) = S(\Lambda[\ket{\psi}\bra{\psi}])$, where $S(\hat{\rho}) = - tr [\rho \log_{2} \rho]$ is the von Neumann entropy. (vi) Additivity: the coherence of a tensor product of quantum states must be additive, $C(\hat{\rho} 	\otimes \hat{\sigma}) = C(\hat{\rho}) + C(\hat{\sigma})$.

Among the many coherence quantifies reported in the literature after the resource theory has arisen \cite{streltsov}, only two were found to fulfill all conditions (i)-(vi). The first is the {\it relative entropy of coherence}, which is defined as \cite{baum}
\begin{equation}
\label{relative coherence}
C_{r}(\hat{\rho}) = S(\hat{\rho}_{diag}) - S(\hat{\rho}),
\end{equation}
where $\hat{\rho}_{diag}$ denotes the state obtained from $\hat{\rho}$ when we delete all off-diagonal entries. In terms of its operational interpretation, it was shown that $C_{r}(\hat{\rho})$ is equal to the {\it distillable coherence} \cite{winter}. The second coherence measure is the {\it coherence of formation}, which was shown to be 
\begin{equation}
\label{coherence of formation}
C_{f}(\hat{\rho}) = min \sum_{n} p_{n} S \left( \ket{\psi_{n}}\bra{\psi_{n}}_{diag} \right) ,
\end{equation}
subject to $\hat{\rho} = \sum_{n} p_{n} \ket{\psi_{n}}\bra{\psi_{n}}$. $C_{f}(\hat{\rho})$ is also known as {\it coherence cost}.

The mathematical properties of $C_{r}(\hat{\rho})$ and $C_{f}(\hat{\rho})$ have been extensively explored in recent works \cite{bu,situ,zhu,zhao}. However, very little has been said about the measurability of these quantities in real experiments. In fact, apart from measuring the contrast of the interference fringes that the quantum system may provide, as suggested by our experimental reasoning \cite{wang}, the usual method to measure coherence is by realizing quantum state tomography to find out $\hat{\rho}$ \cite{zhang}, and then proceed to calculate the coherence measure by means of the appropriate expression. The disadvantage of this procedure is that the state-tomography method, which estimates the quantum state most compatible with a diverse group of measurements, usually provides excessive and redundant information. In this regard, based on the rather simple expressions of the coherence measures, we may argue that it is not necessary to measure the complete information about the system quantum state. In this form, it would be valuable to design experiments to directly measure coherence without appealing to tomographic methods. 

In this work, we introduce an approach to directly measure the relative entropy of coherence $C_{r}(\hat{\rho})$, where by {\it direct} we mean without the need of a complicated set of measurements and calculations usually found in the quantum-state tomography technique. In doing so, we initially consider a direct method for measuring the von Neumann entropy $S(\hat{\rho})$ of an arbitrary quantum state $\hat{\rho}$ in terms of the Shannon entropy of the measurement outcomes obtained in a specific basis. From that, we study how to measure $S(\hat{\rho}_{diag})$, and hence obtain the precise value of $C_{r}(\hat{\rho})$. To illustrate the simplicity and directness of the method, we formulate and discuss two quantum-optical experimental setups aimed to measure the relative entropy of spatial coherence and the coherence of polarization of arbitrary photonic states. The article is organized as follows. In Sec. II we examine the mechanism to directly measure the von Neumann entropy. In Sec. III we explore the method to measure the relative entropy of coherence of an arbitrary state. These results are applied to two important quantum-optical scenarios in Sec. IV. Finally, in Sec. V we discuss our results and summarize our conclusions.

\section{Measuring the von Neumann Entropy} 

In the current section we provide an alternative proof of a theorem that shows that the von Neumann entropy of an arbitrary quantum state is equal to the minimum Shannon entropy of the probability distribution of the measurement outcomes when all possible measurement bases are considered. This is a result that remarkably links two important concepts of the quantum and classical theories of information and that will be essential for our main purpose of directly measuring the coherence of a quantum system.   

One of the cornerstones of the theory of information is the concept of Shannon entropy \cite{shannon,plenio}, which quantifies the information content (unpredictability) of a classical system that can assume $n$ different states with probability $p_{i}$,     
\begin{equation}
\label{shannon}
S=-\sum_{i=1}^{n} p_{i} \log_{2} p_{i}.
\end{equation}
Here, we are aiming to apply this definition to estimate the entropic uncertainty of the probabilistic distribution of outcomes for a quantum measurement of a given observable $\hat{O}$, whose spectral decomposition is $\hat{O}=\sum_{i=1}^{n} e_{i} \ket{e_{i}}\bra{e_{i}}$. We have $e_{i}$ and $\ket{e_{i}}$ representing respectively the $n$ possible eigenvalues and the corresponding eigenstates of the observable. In this case, if the quantum state of the system is given by the density matrix $\hat{\rho}$, the probability to obtain a given eigenvalue in a measurement of $\hat{O}$ is given by \cite{sakurai} 
\begin{equation}
\label{probabilities}
p_{i}=\langle \hat{\Pi}_{i} \rangle =  tr (\hat{\rho} \hat{\Pi}_{i}),
\end{equation}
where $\hat{\Pi}_{i} = \ket{e_{i}}\bra{e_{i}}$ is the projection operator into the eigenstate $\ket{e_{i}}$, and $tr$ denotes the trace.

Having introduced Eqs.~(\ref{shannon}) and~(\ref{probabilities}), we can define the Shannon entropy for the measurement outcomes of $\hat{O}$ as

\begin{equation}
\label{shannon2}
S=-\sum_{i=1}^{n} [tr (\hat{\rho} \hat{\Pi}_{i})]  \log_{2} [tr (\hat{\rho} \hat{\Pi}_{i})]. 
\end{equation}
It is important to call attention to the fact that $n$ represents the dimension of the vector space which describes the observable $\hat{O}$ and the state of the system. If we write the density matrix in the diagonal form, $\hat{\rho}=\sum_{j=1}^{n} \eta_{j} \ket{\eta_{j}}\bra{\eta_{j}}$, the measurement probabilities become $p_{i} = tr (\hat{\rho} \hat{\Pi}_{i}) = tr [(\sum_{j=1}^{n} \eta_{j} \ket{\eta_{j}}\bra{\eta_{j}}) \ket{e_{i}}\bra{e_{i}}]$. This expression can be further simplified by using the trace operation, $p_{i} = \sum_{j=1}^{n} \eta_{j} \braket{e_{i}|\eta_{j}}\braket{\eta_{j}|e_{i}} = \sum_{j=1}^{n} \eta_{j}|\braket{e_{i}|\eta_{j}}|^{2}$. Therefore, by substituting this result into Eq.~(\ref{shannon2}), with $a_{ij} = |\braket{e_{i}|\eta_{j}}|^{2}$ representing the probability of obtaining the eigenvalue $e_{i}$ in a measurement of $\hat{O}$, given that the system was previously in the state $\ket{\eta_{j}}$, we have that

\begin{equation}
\label{shannon3}
S=-\sum_{i=1}^{n} \left(\sum_{j=1}^{n} \eta_{j} a_{ij} \right) \log_{2} \left(\sum_{j=1}^{n} \eta_{j} a_{ij} \right).
\end{equation}
  
At this point we want to ask the following question: which observable of dimension $n$ minimizes the Shannon entropy of Eq.~(\ref{shannon3})? From a similar perspective, we may also ask which set of probabilities $a_{ij}$ minimizes the unpredictability of the measurement outcomes? Before answering these questions, we must note that the probabilities $a_{ij}$ are subject to the following pair of constraints:
\begin{equation}
\label{constraint}
\sum_{ij=1}^{n} a_{ij} = n 
\end{equation}
and
 \begin{equation}
 \label{constraint2}
\sum_{ij=1}^{n} \eta_{j} a_{ij} = 1, 
\end{equation}
both of which reflect the normalization of the quantum state $\hat{\rho}$. 

In order to find the set of probabilities that extremizes the entropy of Eq.~(\ref{shannon3}), given the constraints of Eqs.~(\ref{constraint}) and~(\ref{constraint2}), we shall apply the Lagrange multiplier method. In doing so, we must extremize the Lagrangean function 
\begin{equation}
\label{lagrangean}
L= S + \lambda \left(1 - \sum_{ij=1}^{n} \eta_{j} a_{ij} \right) + \mu \left( n - \sum_{ij=1}^{n}  a_{ij} \right), 
\end{equation}
where $\lambda$ and $\mu$ are the Lagrange multipliers, by requiring that $\partial L /\partial a_{ij} = 0$ for all possible $a_{ij}$. This analysis provides the following result
\begin{equation}
\label{lagrangean2}
\frac{\partial L}{\partial a_{ij}}= - \left[ \eta_{j} \log_{2} \left( \sum_{j=1}^{n} \eta_{j} a_{ij} \right) + \eta_{j}\right] - \lambda \eta_{j} - \mu = 0, 
\end{equation}
which yields
\begin{equation}
\label{lagrangean3}
p_{i}=tr (\hat{\rho} \hat{\Pi}_{i}) = \sum_{j=1}^{n} \eta_{j} a_{ij} = \exp[-1-\lambda - \mu/\eta_{j}].
\end{equation}
This probability distribution is independent of the index $i$, which means that the probabilities $p_{i}$ are all the same. Then, given the constraint of Eq.~(\ref{constraint2}), the probability distribution that extremizes the Shannon entropy of Eq.~(\ref{shannon3}) is simply $p_{i}=1/n$. Note that, instead of minimizing the entropy, this distribution maximizes it. However, the fact that the Lagrange multiplier method only provided a maximum for the entropy means that the probability distribution which minimizes it can be found at the border of the domain of $S$.

The border of the domain of the Shannon entropy is reached when we maximize the irregularity of the probability distribution $p_{i} = \sum_{j=1}^{n} \eta_{j} a_{ij}$. To do that, we can assume, without loss of generality, that $\eta_{1} \geq \eta_{2} \geq...\geq \eta_{n}$. In this case, let us first maximize $p_{1} = \eta_{1} a_{11}+\eta_{2} a_{12}+...+\eta_{n} a_{1n}$. From this expression, since $\eta_{1}$ is the largest population fraction of the density matrix $\hat{\rho}$, it is obvious that $p_{1}$ is maximum when $a_{11} = 1$, and $a_{1j} = 0$ for $j \neq 1$. It signifies that $|\braket{e_{1}|\eta_{1}}|^{2}=1$, which, by assuming the inner product as real for simplicity, gives us that $\ket{e_{1}}=\ket{\eta_{1}}$. Provided this result, keeping in mind that $\braket{e_{k}|e_{l}}= \delta_{kl}$, and $a_{21}=0$ (because $a_{11}=1$), we have that $p_{2}=\eta_{1} a_{21}+\eta_{2} a_{22}+...+\eta_{n} a_{2n}$ is maximum when $a_{22} = 1$, and $a_{2j}=0$ for $j \neq 2$. By following this reasoning of maximization of all $p_{i}$ up to $p_{n}$, the general result is that the Shannon entropy of Eq.~(\ref{shannon3}) is a minimum when $a_{ij} = \delta_{ij}$, where $\delta_{ij}$ is the Kronecker delta. It is easy to verify that this result is also valid even if the spectrum of $\hat{\rho}$ is degenerate, i.e., if there are some $\eta_{k} = \eta_{l}$ for $k \neq l$.    

Quite remarkably, we can see that if the result $a_{ij} = \delta_{ij}$ is used in Eq.~(\ref{shannon3}), we find that the minimum value of the Shannon entropy of the outcomes of a quantum measurement, taken over all possible bases $\{\ket{e_{i}}\}$, is given by     
\begin{equation}
\label{neumann}
S=-\sum_{i=1}^{n} \eta_{i} \log_{2} \eta_{i},
\end{equation}
which is exactly the von Neumann entropy of the state $\hat{\rho}$, i.e., $S = -tr(\hat{\rho} \log_{2} \hat{\rho})$. This allows us to state that: the von Neumann entropy of a general quantum state pertaining to a Hilbert space $\mathcal{H}$ is equal to the minimum Shannon entropy of the measurement outcomes taken over all possible measurement bases of $\mathcal{H}$. Different derivations of this theorem can be found in Refs. \cite{nielsen,wilde}.

As a simple example of application of this result, let us consider the case of an ensemble of spin-1/2 particles described by the density operator $\hat{\rho}$. For us to directly measure the von Neumann entropy of this ensemble, we send all particles through a Stern-Gerlach apparatus, whose alignment of the magnetic field is free to rotate in all directions. In this case, according to the discussion above, the von Neumann entropy is equal to the minimum Shannon entropy of the probability distribution of spin-up and spin-down particles taking over all possible directions of the magnetic field. This configuration is achieved for the most unbalanced probability distribution. Let us consider the limiting cases of pure and mixed ensembles in this situation. The former provides a totally unbalanced distribution of spin up and down particles for a given direction, which means that $S(\hat{\rho})=0$, as expected. The latter, on the other hand, always exhibits a uniform distribution of both types, independent of the magnetic field alignment, therefore, $S(\hat{\rho})=\log_{2}2 = 1$ bit. For partially mixed states, one has first to find the most unbalanced distribution, and then associate the von Neumann entropy to the corresponding Shannon entropy.          

In addition to being recognized as a proper quantifier of disorder \cite{neumann,sakurai}, the von Neumann entropy also represents the smallest amount of information required to fabricate a given state $\hat{\rho}$ \cite{plenio}. To some extent, this last point alludes to the result of the theorem shown above concerning the measurability of the quantum state entropy. Moreover, the importance of accessing this quantity experimentally is relevant in many aspects of quantum information science as, for example, the quantification of entanglement in pure bipartite state systems through the estimation of the entanglement entropy \cite{horodecki}, and the description of the coherence properties of quantum systems as a physical resource in terms of the relative entropy of coherence \cite{streltsov,baum,winter}, which will be our focus in the next section. In closing this section, it is worthwhile mentioning that the method of measuring the von Neumann entropy discussed here is direct in the sense that we do not need to learn the quantum state of the system. That is to say that quantum-state tomography is unnecessary. Yet, one could argue that the measurement cost of realizing the present method is similar to that of tomographic methods, once all measurement bases should be tested in order to find the outcome distribution that minimizes the Shannon entropy. However, in Sec. IV we will show that, because of the invariance of $S(\hat{\rho})$ upon unitary transformations acting on $\hat{\rho}$, we can greatly reduce the set of measurement bases possibilities to be tested, and hence the complexity of an experimental realization.               

\section{Measuring the relative entropy of coherence} 

Now that we have explained the scheme to directly access the von Neumann entropy of a quantum state $ S(\hat{\rho})$ in a real experiment, based on the definition of the relative entropy of coherence of Eq.~(\ref{relative coherence}), an auxiliary approach to directly measure the von Neumann entropy of the dephased state $\hat{\rho}_{diag}$ needs to be presented. Here, we have to point out that, instead of $S(\hat{\rho})$, $S(\hat{\rho}_{diag})$ is a basis dependent quantity, as well as the relative entropy of coherence \cite{baum}. Let us write the complete density operator in a given basis $\{\ket{\rho_{i}}\}$ in the following form           
\begin{equation}
\label{densop}
\hat{\rho}=\sum_{ij=1}^{n} \rho_{ij} \ket{\rho_{i}} \bra{\rho_{j}}.
\end{equation}
Then, the dephased density operator is given by
\begin{equation}
\label{densop2}
\hat{\rho}_{diag}=\sum_{i=1}^{n} \rho_{ii} \ket{\rho_{i}} \bra{\rho_{i}},
\end{equation}
yielding the following von Neumann entropy:
\begin{equation}
\label{neumann2}
S(\hat{\rho}_{diag}) = -\sum_{i=1}^{n} \rho_{ii} \log_{2} \rho_{ii}.
\end{equation}
The numbers $\rho_{ii}$ are the diagonal terms of the density matrix in Eq.~(\ref{densop}), which physically represent the probabilities of measuring the system initially prepared in the state $\hat{\rho}$ in the pure states $\ket{\rho_{i}}$, given that a measurement in the basis $\{\ket{\rho_{i}}\}$ was realized.

It is clear from Eq.~(\ref{neumann2}) that it matches the Shannon entropy of the measurement outcomes in the basis $\{\ket{\rho_{i}}\}$. However, it is not difficult to recognize which measurement basis is this. It is precisely the basis in which we want to find out if the state $\hat{\rho}$ possesses coherence. In this form, this basis is chosen according to the context in which the experimentalist desires to learn about the coherence properties of the system. Now that we understand the recipes for measuring both $S(\hat{\rho}_{diag})$ and $S(\hat{\rho})$, the relative entropy of coherence $C_{r}(\hat{\rho})$ can be obtained from the difference between these two quantities, according to the definition of Eq.~(\ref{relative coherence}). In the next section we provide two insightful quantum-optical examples that show how we can apply our predictions to measure the relative entropy of coherence of two beams of light (or the same beam at two separated points), and similarly for the quantum coherence of the polarization state of a single beam in any desired basis.       

\section{Measurability of the relative entropy of coherence in quantum-optical systems} 

It is valuable at this stage to discuss real experiments in which our proposal of directly measuring the relative entropy of coherence can be implemented. The proposals to be presented here will be focused in estimating $C_{r}(\hat{\rho})$ with respect to the spatial coherence between the photons received from two light sources, and in a similar fashion with respect to the coherence properties of the polarization of a single beam of light in a basis of interest.  

\subsection{Experimental proposal for measuring $C_{r}(\hat{\rho})$ to quantify optical spatial coherence}

In quantifying the spatial coherence of two light beams (or the same beam detected at two different points in space) through the measurement of $C_{r}(\hat{\rho})$, we propose the scheme shown in Fig. 1. First, let us focus on the measurement of the von Neumann entropy $S(\hat{\rho})$ of the ensemble of photons with respect to the path taken before detection. In this case, the experimental apparatus consists in an ensemble of photons emitted from the sources $Q_{0}$ and $Q_{1}$, whose quantum state we denote by $\ket{0}$ and $\ket{1}$, respectively. For now, we consider only the case in which the photons have the same probability of taking both paths. i.e., the two sources have the same intensity. By the end of this section we relax this condition. One phase shifter is placed in the lower path with the aim to impart a controllable relative phase $\phi$ between the two path states. After that, the two photon beams are superposed by a 50:50 lossless beam splitter, BS, before being collected by the photo-detectors $D_{0}$ and $D_{1}$. Recall that, for measuring $S(\hat{\rho})$ according to the discussion of Sec. II, we need to measure the path taken by the photons in all possible measurement bases. Since there are two possible paths, in principle we would have to make measurements in all bases $\{\ket{e_{1}},\ket{e_{2}}\}$, where 
\begin{equation}
\label{basis1}
\ket{e_{1}(\theta,\phi)}= \cos\left( \frac{\theta}{2}\right) \ket{0} + e^{i\phi} \sin\left( \frac{\theta}{2}\right) \ket{1}
\end{equation}
and 
\begin{equation}
\label{basis2}
\ket{e_{2}(\theta,\phi)}= \sin\left( \frac{\theta}{2}\right) \ket{0} - e^{i\phi} \cos\left( \frac{\theta}{2}\right) \ket{1}.
\end{equation}

These measurement eigenstates are geometrically described by diametrically opposed points on the surface of the Bloch sphere of Fig. 2, which depend on the polar and azimuthal angles, $\theta$ and $\phi$. However, since we are treating the case in which the probabilities of measuring the photon in the states $\ket{0}$ and $\ket{1}$ are $1/2$, the state of the photon necessarily lies on the equatorial plane of the Bloch sphere, $\theta= \pi/2$. In this case, for us to find the measurement basis which minimizes the Shannon entropy of the outcomes, which corresponds to the von Neumann entropy, we have to look for the measurement eigenstates which lie on the equatorial line of the Bloch sphere. If this is realized, by varying the parameter $\phi$ of the eigenstates, at some point they will necessarily coincide with the states $\ket{\eta}_{j}$ that compose the spectral decomposition of the density matrix of the photon state, $\hat{\rho}=\sum_{j=1}^{2} \eta_{j} \ket{\eta_{j}}\bra{\eta_{j}}$. As discussed in Sec. II, at this point one encounters the most unbalanced measurement outcome distribution for all possible values of $\phi$, which is easy to find experimentally. In what follows, we give more details concerning the experiment.
\begin{figure}[htb]
\begin{center}
\includegraphics[height=1.25in]{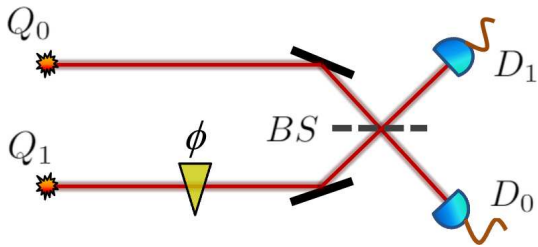}
\end{center}
\label{F1}
\caption{(color online). Scheme of the apparatus to measure the relative entropy of (spatial) coherence of the photons that are emitted by the sources $Q_{0}$ and $Q_{1}$ with equal probability. The setup requires only linear optical devices: a phase shifter in the lower path that imparts a relative phase $\phi$ between the two paths, a $50:50$ beam splitter $BS$ and two photo-detectors $D_{0}$ and $D_{1}$. Two mirrors (black traces) are used to orient the beams towards the input ports of $BS$.}
\end{figure}

In the theory of optical quantum information processing, the 50:50 BS is shown to perform the following operation between the input and output modes, $\hat{B}=1/\sqrt{2}(\ket{0}\bra{0}-\ket{0}\bra{1}+\ket{1}\bra{0}+\ket{1}\bra{1})$, and the phase shifter the operation $\hat{S}_{\phi}=\ket{0}\bra{0}+e^{i \phi} \ket{1}\bra{1}$ (See Refs. \cite{nielsen,fox,kok}). With the configuration of Fig. 1, the detector $D_{0}$ certainly registers a click when the initial state is $\ket{D_{0}} = 1/\sqrt{2}(\ket{0}-e^{-i \phi} \ket{1})$, and the same is valid for the detector $D_{1}$ with the initial state $\ket{D_{1}} = 1/\sqrt{2}(\ket{0}+e^{-i \phi} \ket{1})$. As a result, we have that the apparatus of Fig. 1 realizes a measurement of the path taken by the photons in the basis $\{\ket{D_{0}},\ket{D_{1}}\}$. It signifies that by continuously varying the relative phase $\phi$ between the two paths, we can make all possible path measurements whose pair of eigenstates lie at diametrically opposed points on the equatorial (dashed) line of the Bloch sphere. 

\begin{figure}[htb]
\begin{center}
\includegraphics[height=2.55in]{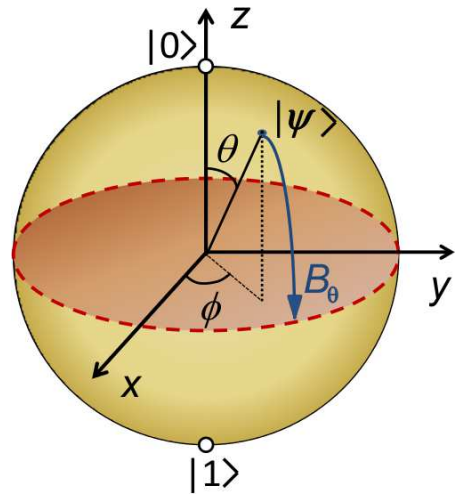}
\end{center}
\label{F1}
\caption{(color online). Bloch sphere representation of the which-path states of the photons coming from the sources $Q_{0}$ and $Q_{1}$ in the apparatuses of Figs. 1 and 3. The rotation operation $\hat{B}_{\theta}$ that takes place by using an adequate beam splitter, as shown in Fig. 3, aims to move an arbitrary vector state to the equatorial plane, which is limited by the dashed circumference.}   \end{figure}

As a matter of fact, if we are interested in measuring the spatial coherence of the light sources in the basis $\{\ket{0},\ket{1}\}$, we still have to explain how to measure $S(\rho_{diag})$ in this basis. The obvious answer is by removing the phase shifter and $BS$, and registering the {\it normalized} intensities of the two beams, $I_{0}$ and $I_{1}$, for the photons coming from Q$_{0}$ and Q$_{1}$, respectively. In this case, $S(\rho_{diag})$ is precisely the Shannon entropy of the outcomes of the measurement in this basis, which is simply given by $S(\rho_{diag}) = -\sum_{k=0,1}I_{k} \log_{2} I_{k}$. So far, we are considering only the cases in which $I_{0} = I_{1} = 1/2$, then we have simply that $S(\hat{\rho}_{diag}) = \log_{2} 2 = 1$ bit. Therefore, having measured both $S(\hat{\rho})$  and $S(\hat{\rho}_{diag})$, one can calculate the relative entropy of coherence by means of Eq.~(\ref{relative coherence}). 

We now present the experimental method to measure $S(\hat{\rho})$ and $C_{r}(\hat{\rho})$, in the general case when $I_{0} \neq I_{1}$. The scheme is analogous to that when $I_{0} = I_{1}$ of Fig. 1, but with a previous stage that realizes a unitary operation $\hat{U}$ on the initial state of the photons. This operation consists in a rotation of the state to the equatorial plane. As seen above, if the state ends up on this plane, it means that $I_{0}=I_{1}$ so that the setup of Fig. 1 can be then used to determine $S(\hat{\rho})$. Note that the von Neumann entropy of the quantum state is unchanged under unitary transformations, $S(\hat{\rho}) = S(\hat{U}\hat{\rho}\hat{U}^{\dagger})$. In this form, by measuring the entropy of the rotated state $S(\hat{U}\hat{\rho}\hat{U}^{\dagger})$, which is realized with the setup of Fig. 1, one measures the entropy of the original state $S(\hat{\rho})$. In this case, to conclude the protocol of measuring the von Neumann entropy of an arbitrary photonic path state, it suffices to show how to rotate a general state towards the equatorial plane of the Bloch sphere in Fig. 2. The apparatus is shown in Fig. 3.    
\begin{figure}[htb]
\begin{center}
\includegraphics[height=1.3in]{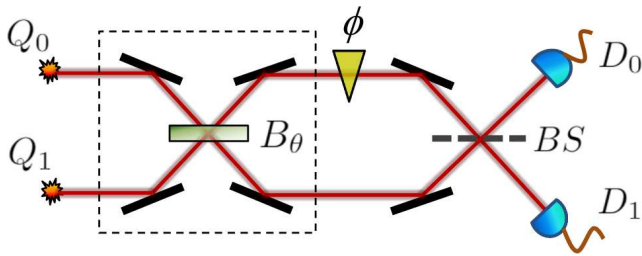}
\end{center}
\label{F1}
\caption{(color online). Experimental scheme to measure the relative entropy of coherence of arbitrary which-path states. The only difference between this setup and that of Fig. 1 is the primary stage, involved by the dashed lines, which executes the state rotation as shown in Fig. 2.}
\end{figure}

As can be seen, the experimental setup of Fig. 3 is identical to that of Fig. 1, except for the initial stage, involved by the dashed lines, which is intended to rotate the path states of the photons to the equatorial plane of the Bloch sphere. The central element at this stage is the beam splitter $B_{\theta}$ that receives the beams coming from the sources $Q_{0}$ and $Q_{1}$, with different intensities, and creates output modes with the same intensity. In the appendix of this work we show how to choose an adequate beam splitter for this task. After this point, the state of the system is certainly lying on the equatorial plane, and the rest of the experiment is described by the same protocol of Fig. 1, which directly provides the elements to calculate $S(\hat{\rho})$. At last, as already mentioned, the entropy $S(\hat{\rho}_{diag})$ of an arbitrary photonic path state is simply given by  
$S(\hat{\rho}_{diag}) = -\sum_{k=0,1}I_{k} \log_{2} I_{k}$, where $I_{0}$ and $I_{1}$ are the normalized intensities of the sources $Q_{0}$ and $Q_{1}$, respectively. Thus, given the measurement descriptions of $S(\hat{\rho}_{diag})$ and $S(\hat{\rho})$, the relative entropy of coherence for an arbitrary photonic which-path state can be realized with the apparatus of Fig. 3.

\subsection{Experimental proposal for measuring $C_{r}(\hat{\rho})$ to quantify the coherence of polarization}

Here we propose an experimental setup to measure the relative entropy of coherence of polarization in a given basis. Let us first focus on the measurement of $S(\hat{\rho})$, whose scheme is shown in Fig. 4. It consists of a beam of photons coming from a source $Q$ passing through a half-wave plate ($HWP_{1}$) before entering a Mach-Zehnder like interferometer. The aim of $HWP_{1}$ is to rotate the polarization state of the photons around the $y$ axis of the Poincar\'e sphere towards the equatorial plane \cite{kok,fox,hecht}, as shown in Fig 5. Specifically, if the optical axis of this plate makes an angle $\alpha$ with the horizontal direction, it causes a rotation of $2 \alpha$ in the state vector around the $y$ axis. After that, the beam passes through a polarizing beam splitter ($PBS$) that transmits horizontal and reflects vertical polarization states, $\ket{H}$ and $\ket{V}$ respectively. Since the polarization state will lie on the equatorial plane of the Poincar\'e sphere after $HWP_{1}$, the beams reflected and transmitted by $PBS$ will have the same intensity. The vertically polarized beam, which takes the superior arm of the interferometer, passes through a second half-wave plate $HWP_{2}$ that transforms the polarization state into the state $\ket{H}$ before passing through a phase shifter that imparts a controllable relative phase $\phi$ between the two beams in the interferometer. The last stage of the scheme is composed by a 50:50 beam splitter $BS$ and two photo-detectors localized at the output ports, $D_{0}$ and $D_{1}.$     
\begin{figure}[htb]
\begin{center}
\includegraphics[height=1.95in]{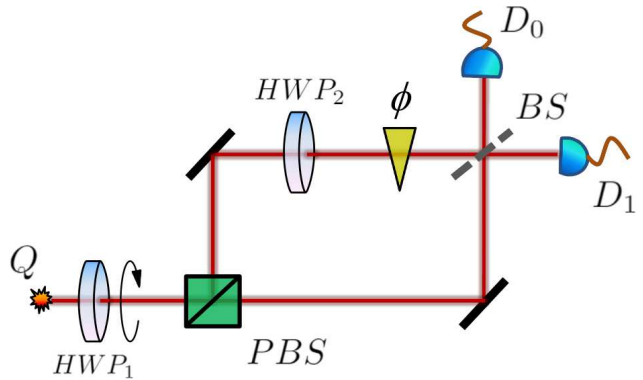}
\end{center}
\label{F1}
\caption{(color online). Experimental setup to measure the relative entropy of coherence of arbitrary polarization states, which requires only linear-optical devices. In-depth explanation about its operation can be found in the main text.}   
\end{figure}

The role of $HWP_{2}$ is to rotate the polarization for the beams to interfere at $BS$ prior to detection. Then, with this configuration, $D_{0}$ certainly register a click if the polarization state before $PBS$ is $\ket{D_{0}} = 1/ \sqrt{2} (\ket{H}- e^{-i \phi}\ket{V})$. The same argument applies to $D_{1}$ if the state before $PBS$ is $\ket{D_{1}} = 1/ \sqrt{2} (\ket{H} + e^{-i \phi}\ket{V})$. In this case, by continuously varying the phase shift $\phi$ from $0$ to $2 \pi$, we realize all measurements whose basis states lie on the equatorial line of the Poincar\'e sphere. Observe that, once the polarization state of the photons necessarily lie on the equatorial plane after $HWP_{1}$, the minimum Shannon entropy of the intensity distribution at $D_{0}$ and $D_{1}$, which corresponds to the von Neumann entropy of the original polarization state, can be certainly found for a given value of $\phi$. As already pointed out, this minimum Shannon entropy is reached for the most unbalanced intensity distribution registered by the detectors. It is worthwhile to emphasize that the rotation of the polarization state caused by $HWP_{1}$, which is an unitary operation, does not alter the von Neumann entropy of the original quantum state coming from $Q$. In this form, the experimental setup of Fig. 4 does provide direct access to $S(\hat{\rho})$.               
\begin{figure}[htb]
\begin{center}
\includegraphics[height=2.55in]{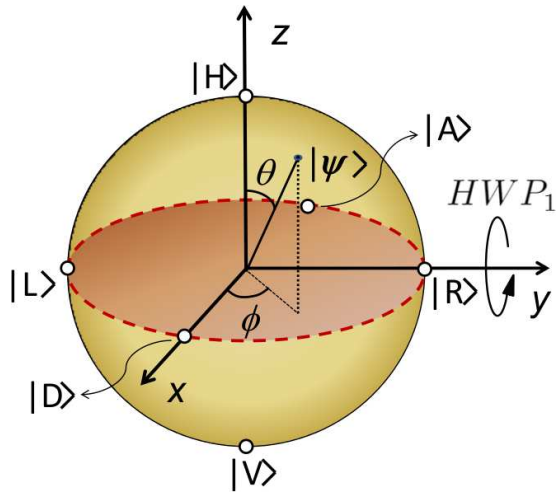}
\end{center}
\label{F1}
\caption{(color online). Poincar\'e sphere representing all possible polarization states. The effect of $HWP_{1}$ in the scheme of Fig. 4 is to rotate the state around the $y$-axis towards the equatorial plane. The small open circles represent horizontal and vertical polarization states, $\ket{H}$ and $\ket{V}$; diagonal and anti-diagonal polarization states, $\ket{D}$ and $\ket{A}$; and left circular and right circular polarization states, $\ket{L}$ and $\ket{R}$.}   
\end{figure}

Since we have learned how to measure $S(\hat{\rho})$, for us to calculate $C_{r}(\hat{\rho})$ for the polarization in the basis $\{ \ket{H}, \ket{V}\}$, we need to measure $S(\hat{\rho}_{diag})$ in this basis. It is trivial to do this with the apparatus of Fig. 4. We only have to remove $HWP_{1}$, $HWP_{2}$, $BS$ and the phase shifter, and  measure the normalized intensities $I_{0}$ and $I_{1}$ at the detectors. In this case, we have that $S(\hat{\rho}_{diag}) = -\sum_{k=0,1}I_{k} \log_{2} I_{k}$. Thus, since we possess sufficient knowledge about $S(\hat{\rho})$ and $S(\hat{\rho}_{diag})$, we can calculate $C_{r}(\hat{\rho})$ by using Eq.(\ref{relative coherence}). It is worthwhile mentioning that the setup of Fig. 4 can be used to measure the relative entropy of coherence for the polarization in any desired basis. In fact, for a different basis the only quantity that changes is $S(\hat{\rho}_{diag})$. Thus, we just use the same configuration of Fig. 4 (without $HWP_{1}$, $HWP_{2}$, $BS$ and the phase shifter) and place a quarter-wave plate and a half-wave plate before PBS to choose the desired basis in which we want to learn about the coherence properties. The orientation of the optical axis of these plates allows one to select any basis states corresponding to diametrically opposed points of the Poicar\'e sphere \cite{nielsen, kok}.

\section{conclusion} 

In this work, we have proposed a method to directly measure the von Neumann entropy $S(\hat{\rho})$ and the relative entropy of coherence $C_{r}(\hat{\rho})$ of arbitrary quantum states. By direct we mean that it is not necessary to learn the state $\hat{\rho}$ through, for example, quantum-state tomography. To develop these methods, we used the fact that the von Neumann entropy of a quantum state is equal to the minimum Shannon entropy of the measurement outcomes taken over all possible measurement bases. In fact, we presented an alternative derivation for this result. Also, we have applied the concept of Shannon entropy to access the von Neumann entropy of the dephased state $S(\hat{\rho}_{diag})$. To support our proposal, we described two quantum-optical applications aimed to have a direct experimental access to the relative entropy of spatial and polarization coherence. The experimental proposals were found to give the exact results for $C_{r}(\hat{\rho})$ in both situations. Remarkably, due to the fact that $S(\hat{\rho})$ is invariant upon unitary transformations on $\hat{\rho}$, we showed that the set of measurement bases to test the minimization of the Shannon entropy can be extremely reduced, which makes the method considerably efficient.  

The experiments proposed here involve the which-path information in a two-way interferometer and the polarization state of photons, which could be treated with the general formalism of two-state quantum systems. Therefore, we see no obstacle to extend this approach to the study of the coherence properties of spin-1/2 particles as, for example, in NMR quantum computing systems that can be manipulated with externally applied electromagnetic fields \cite{vand}, or (effective) two-level atoms in cavity quantum electrodynamics (QED) experiments \cite{raim}. Overall, given the importance of the concepts of von Neumann entropy and relative entropy of coherence to the field of quantum information, we believe that our direct strategy to experimentally access them may provide a valuable alternative to quantum state tomography.

\begin{acknowledgements}
The author acknowledges financial support from the Brazilian funding agencies CAPES/Finance Code 001, and CNPq, Grant Number 309292/2016-6.
\end{acknowledgements}

\section*{APPENDIX: CHOOSING THE ADEQUATE BEAM SPLITTER $B_{\theta}$ FOR THE OPTICAL CIRCUIT OF FIG. 3}
\renewcommand{\theequation}{A-\arabic{equation}}
\setcounter{equation}{0}

Here we shall describe how to choose the appropriate beam splitter $\hat{B}_{\theta}$ at the first stage of the setup of Fig. 3. As discussed in Sec. IV, the role of $B_{\theta}$ is to rotate the quantum state towards the equatorial plane of the Bloch sphere, as shown in Fig. 2. Experimentally, it signifies that the two output modes of the beam splitter must have the same intensity. From the theory of linear optical devices, the unitary matrix associated with a generic beam splitter $B_{\theta,\phi}$ is \cite{kok}
\begin{equation}
\label{BS1}
\hat{B}_{\theta,\phi} = 
\begin{pmatrix}
cos(\theta) & -e^{i\phi} sin (\theta) \\
e^{-i\phi} sin (\theta) & cos(\theta)
  \end{pmatrix}.
\end{equation} 
The reflection and transmission coefficients of
the beam splitter are $R= \sin^{2}(\theta)$ and $T=1 -R=\cos^{2}(\theta)$, respectively. The
relative phase shift $	\pm e^{\mp i\phi}$ guarantees that the transformation
is unitary. 

Without loss of generality, we choose here $\phi = 0$, such that the matrix of the beam splitter 
of Fig. 3 is the rotation matrix
\begin{equation}
\label{BS2}
\hat{B}_{\theta} = 
\begin{pmatrix}
cos(\theta) & - sin (\theta) \\
sin (\theta) & cos(\theta)
  \end{pmatrix}.
\end{equation} 
Thus, after passing through the first stage of the circuit in Fig. 3 (involved by the dashed lines), an initial quantum state $\hat{\rho}$ will be transformed into the state $\hat{\rho}\ensuremath{'}= \hat{B}_{\theta} \hat{\rho} \hat{B}^{\dagger}_{\theta}$. In this form, for us to obtain an output state $\hat{\rho}\ensuremath{'}$ that lies on the equatorial plane, we just need to choose a beam splitter with an appropriate angle $\theta$. That is to say that $R$ and $T$ are such that the output intensities are the same. In what follows, we give two numerical examples which support the validity of the complete scheme of Fig. 3 in determining the relative entropy of spatial coherence. We first illustrate with an incoherent state and later with a partially coherent state.     
\\\\
\noindent 
{\it Incoherent State}: Let us consider the case of the incoherent state
\begin{equation}
\label{state1}
\hat{\rho} = 3/4 \ket{0}\bra{0}+1/4 \ket{1}\bra{1}
\end{equation}
coming from the sources $Q_{0}$ and $Q_{1}$ in Fig. 3. After passing through a beam splitter $B_{\theta}$, with $\theta = 0.785$ rad, and the phase shifter, the state becomes $\hat{\rho}\ensuremath{''}= \hat{S}_{\phi}\hat{B}_{\theta} \hat{\rho} \hat{B}^{\dagger}_{\theta} \hat{S}^{\dagger}_{\phi}$, which is given by
\begin{equation}
\label{state11}
\hat{\rho}\ensuremath{''} = \begin{pmatrix}
1/2 & e^{-i \phi}/4 \\
e^{i \phi}/4 & 1/2
  \end{pmatrix}.
\end{equation}
This is the state before $BS$. Thus, the probabilities to measure the photon at $D_{0}$ and $D_{1}$ are given respectively by 
\begin{equation}
\label{p0}
P_{0} = tr (\hat{\rho}\ensuremath{''} \hat{\Pi}_{0}) = \frac{1}{2} - \frac{1}{4} \cos(2 \phi)
\end{equation}
and 
\begin{equation}
\label{p1}
P_{1} = tr (\hat{\rho}\ensuremath{''} \hat{\Pi}_{1}) = \frac{1}{2} + \frac{1}{4} \cos(2 \phi),
\end{equation}
where $\hat{\Pi}_{0}= \ket{D_{0}}\bra{D_{0}}$ and $\hat{\Pi}_{1}= \ket{D_{1}}\bra{D_{1}}$. From Eqs.~(\ref{p0}) and~(\ref{p1}), we see that the most unbalanced distribution is when $\phi = 0$, i.e., $P_{0} = 1/4$ and $P_{1} = 3/4$, which according to the result of Sec. II gives us that the von Neumann entropy of the state is $S(\hat{\rho}) = -1/4 \log_{2} (1/4) - 3/4 \log_{2} (3/4) \approx 0.811$ bit. It is easy to see that $S(\hat{\rho}_{diag})$ is obtained from the Shannon entropy of the normalized intensities of $Q_{0}$ and $Q_{1}$ that are respectively $I_{0} = 3/4$ and $I_{1} = 1/4$, which yields $S(\hat{\rho}_{diag})=S(\hat{\rho}) \approx 0.811$ bit. Therefore, from Eq.~(\ref{relative coherence}) we obtain that $C_{r}(\hat{\rho}) = 0$; the expected value for an incoherent state. \\\\

\noindent 
{\it Partially Coherent State}: Now we consider the initial state
\begin{equation}
\label{state2}
\hat{\rho} = 3/4 \ket{+}\bra{+}+1/4 \ket{0}\bra{0},
\end{equation}
with $\ket{+} = 1/\sqrt{2} (\ket{0}+\ket{1})$. After the $\hat{B}_{\theta}$ (with $\theta \approx 0.161$ rad) and $\hat{S}_{\phi}$ operations, we obtain $\hat{\rho}\ensuremath{''}= \hat{S}_{\phi}\hat{B}_{\theta} \hat{\rho} \hat{B}^{\dagger}_{\theta} \hat{S}^{\dagger}_{\phi}$, which can be shown to be
\begin{equation}
\label{state22}
\hat{\rho}\ensuremath{''} \approx \begin{pmatrix}
1/2 & 0.395 e^{-i \phi} \\
0.395 e^{i \phi} & 1/2
  \end{pmatrix}.
\end{equation}
This is the state before $BS$. Then, the probabilities to detect the photons at $D_{0}$ and $D_{1}$ are given by 
\begin{equation}
\label{p00}
P_{0} = tr (\hat{\rho}\ensuremath{''} \hat{\Pi}_{0}) = \frac{1}{2} - 0.395 \cos(2 \phi)
\end{equation}
and 
\begin{equation}
\label{p11}
P_{1} = tr (\hat{\rho}\ensuremath{''} \hat{\Pi}_{1}) = \frac{1}{2} + 0.395 \cos(2 \phi),
\end{equation}
respectively. These results show that the most unbalanced distribution occurs for $\phi = 0$, i.e., $P_{0} \approx 0.105$ and $P_{1} \approx 0.895$. In this form, the von Neumann entropy of the state of Eq.~(\ref{state2}) is $S(\hat{\rho}) \approx -0.105 \log_{2} (0.105) - 0.895 \log_{2} (0.895) \approx 0.485$ bit. This result is in accordance with the calculated value $S(\hat{\rho}) = -tr[\hat{\rho}\log_{2}\hat{\rho}]$ for this state. Again, we have that $S(\hat{\rho}_{diag})$ can be found directly from the Shannon entropy of the normalized intensities of $Q_{0}$ and $Q_{1}$, which are obtained from the state of Eq.~(\ref{state2}) as $I_{0} = tr(\hat{\rho}\ket{0}\bra{0})=5/8$ and $I_{1} = tr(\hat{\rho}\ket{1}\bra{1})=3/8$. Then, we find $S(\hat{\rho}_{diag}) \approx 0.954$ bit. Finally, from Eq.~(\ref{relative coherence}) we obtain that $C_{r}(\hat{\rho}) \approx 0.469$ cobit; a result that indicates partial coherence. The special case of a {\it maximally coherent state} as, for example, the pure state $\hat{\rho}=\ket{+}\bra{+}$ can be analyzed in a simple way with the setup of Fig. 1.


\end{document}